\documentclass[prl,twocolumn]{revtex4}
\usepackage{amsmath,amssymb}
\usepackage{amsthm}
\usepackage{epsfig}
\usepackage{xcolor}
\usepackage{fixltx2e}
\usepackage[normalem]{ulem}
\usepackage{siunitx}

\usepackage{xr}
\newcommand {\pt}{\partial}

\newcommand{\bsf}{\begin{subfigure}} 
\newcommand{\esf}{\end{subfigure}} 

\graphicspath{{eps/}}            

\begin{document}
\title{Manipulation of strongly interacting solitons in optical fiber experiments}

\author{Alexandre Mucci$^{1}$}
\author{Pierre Suret$^{1}$}
\author{François Copie$^{1}$}
\author{Stephane Randoux$^{1}$}
\author{Rustam Mullyadzhanov$^{2,3}$}
\author{Andrey Gelash$^{4}$}
\email[Corresponding author : ]{agelash@gmail.com}
\thanks{Current address: Institute of Physics, Swiss Federal Institute of Technology Lausanne (EPFL), CH-1015 Lausanne, Switzerland.}

\affiliation{$^{1}$University Lille, CNRS, UMR 8523 - PhLAM -  Physique des Lasers Atomes et Mol\'ecules, F-59000 Lille, France}
\affiliation{$^{2}$Novosibirsk State University, Novosibirsk 630090, Russia}
\affiliation{$^{3}$Institute of Thermophysics SB RAS, Novosibirsk 630090, Russia}
\affiliation{$^{4}$Laboratoire Interdisciplinaire Carnot de Bourgogne (ICB), UMR 6303 CNRS-Université Bourgogne Franche-Comté, 21078 Dijon, France}

\begin{abstract}
The model underlying physics of guiding light in single-mode fibers -- the one-dimensional nonlinear Schr\"odinger equation (NLSE), reveals a remarkable balance of the fiber dispersion and nonlinearity, leading to the existence of optical solitons. With the Inverse Scattering Transform (IST) method and its perturbation theory extension one can go beyond single-soliton physics and investigate nonlinear dynamics of complex optical pulses driven by soliton interactions. Here, advancing the IST perturbation theory approach, we introduce the eigenvalue response functions, which provide an intuitively clear way to manipulate individual characteristics of solitons even in the case of their entire overlapping, i.e. very strong interactions. The response functions reveal the spatial sensitivity of the multi-soliton pulse concerning its instantaneous perturbations, allowing one to manipulate the velocities and amplitudes of each soliton. Our experimental setup, based on a recirculating optical fiber loop system and homodyne measurement, enables observation of long-distance NLSE dynamics and complete characterization of optical pulses containing several solitons. Adding a localized phase perturbation on a box-shaped wavefield, we change individual soliton characteristics and can detach solitons selectively from the whole pulse. The detached solitons exhibit velocities very close to the theoretical predictions, thereby demonstrating the efficiency and robustness of the response functions approach.
\end{abstract}
\maketitle
Experimental observation of optical solitons in single-mode fibers reported in 1980 \cite{mollenauer1980experimental} and the following demonstration of their stable propagation over thousands of kilometers maintained by a loss compensation \cite{mollenauer1988demonstration} opened avenues into fundamental studies and engineering applications in the field of nonlinear optics \cite{kivshar2003optical,hasegawa2013optical}. The model underlying physics of guiding light in fibers -- the one-dimensional nonlinear Schr\"odinger equation (NLSE), reveals a remarkable balance of the fiber dispersion and nonlinearity, leading to the effect of soliton shape preservation \cite{hasegawa1973transmissionI,hasegawa1973transmissionII,kivshar2003optical,remoissenet2013waves}. Manipulation of nonlinear light pulses represents a long-standing challenge in fiber optics. Nowadays, thanks to modern ultra-fast lasers, waveshapers, and electro-optic modulators, the reproduction and management of individual solitons in optical fibers become a routine procedure used e.g., in supercontinuum generation \cite{dudley2006supercontinuum,skryabin2010colloquium} or as a tool for exploring various optical phenomena \cite{jang2015temporal,suret2023soliton,englebert2024manipulation}.

To go beyond single-soliton physics, one can rely on a powerful theoretical technique -- the Inverse Scattering Transform (IST) method based on the integrable nature of the NLSE \cite{zakharov1972exact,AblowitzBook1981,NovikovBook1984}. The IST realizes a mapping of nonlinear wavefield dynamics onto linear evolution of the IST spectrum, more rigorously referred to as scattering data. In this sense, the IST procedure and scattering data are nonlinear analogs of the conventional Fourier transform and the Fourier spectra respectively \cite{OsborneBook2010,AblowitzBook1981}. Considering the focusing NLSE, the IST spectrum consists of discrete and continuous components corresponding to solitons and nonlinear dispersive waves. One can find the IST spectrum by solving an auxiliary linear Zakharov-Shabat (ZS) eigenvalue system and vice versa -- obtain the wave field from known IST spectrum. For example, soliton interactions can be described using exact multi-soliton formulas, parameterized by the IST soliton eigenvalues. In general case, the evolution of initial conditions containing non-zero continuous spectrum parts can be expressed via the integral Gelfand-Levitan-Marchenko (GLM) equations. In addition, the well-developed IST perturbation theory allows one to approximate the dynamics of perturbed solitons and generalize the IST approach to nearly integrable models, e.g., the NLSE with third-order dispersion, gain-loss or Raman scattering terms \cite{wabnitz1993suppression,afanasjev1995soliton,okamawari1995analyses,gerdjikov1996asymptotic,KivsharRMP1989,YangBook}.

In this Letter, we consider theoretically and study experimentally nonlinear propagation of perturbed box-shaped light pulses in an optical single-mode fiber system used in the anomalous dispersion regime. The pulses consist of several solitons and a portion of nonlinear dispersive waves, which complicates the straightforward IST analysis because, in this case, the GLM equations can only be solved asymptotically \cite{NovikovBook1984,AblowitzBook1981,jenkins2014semiclassical}. Meanwhile, numerical simulations, always available for the NLSE model, do not give a general description of the problem. To cope with the analysis difficulties we advance the IST perturbation theory and introduce the eigenvalue response functions (RFs) which provide an intuitively clear way to manipulate strongly interacting solitons in nonlinear pulses. The RFs are computed individually for each of the solitons via the corresponding eigenfunction solutions of the ZS system and reveal the spatial sensitivity of the pulse concerning wavefield perturbations. Being linked to the integral describing change of soliton eigenvalues in response to the perturbation, the RFs allow one to manipulate individual soliton velocities and amplitudes perturbing the pulse in a controllable way.

As a proof of concept, we design experiments for a recirculating optical fiber loop system with Raman amplifier and homodyne interferometric tools, enabling the recording of long-distance NLSE spatiotemporal dynamics together with the full characterization of complex optical pulses (phase and amplitude). Adding a localized phase perturbation of the light pulse we manage to change individual soliton characteristics and even detach a desired soliton or a soliton complex from the whole pulse. The detached solitons being measured in the fiber loop system exhibit velocities very close to the theoretical ones, thereby demonstrating the efficiency and robustness of the proposed approach.

The model of light propagation in our experimental setup is the NLSE with effective losses term \cite{kraych2019nonlinear,Kraych2019,suret2023soliton},
\begin{eqnarray}
\label{dim_NLSE}
i \frac{\pt \Psi}{\pt Z} = \frac{\beta_2}{2} \frac{\pt^2 \Psi}{\pt T^2} - \gamma |\Psi|^2 \Psi - i\frac{\alpha}{2}\Psi,
\end{eqnarray}
where $\Psi(Z,T)$ is the complex envelope of the electric field $E(Z,t)=\Psi(Z,t-z/v_g)\,\exp{i(k_0 Z-\omega_0 t)}+\text{c.c}$, that slowly varies in space $Z$ and the retarded time $T=t-Z/v_g$, measured in the frame propagating at the group velocity $v_g=v_g(\omega_0)$. The laboratory frame time is $t$, while $\omega_0$ and $k_0$ represent the carrier wave frequency and wavenumber. At our working wavelength $1550$~nm, the group velocity dispersion coefficient is $\beta_2 = -22 \text{ ps}^2/\text{km}$, the nonlinear Kerr coefficient is $\gamma = 1.3 \text{W}^{-1}\text{km}^{-1}$ and the recirculating fiber loop losses $\alpha =  \qty{4d-7}{m^{-1}}$.

Our theoretical treatment is based on the dimensionless form of the NLSE,
\begin{eqnarray}
\label{NLSE}
i \psi_z + \frac12 \psi_{\tau\tau} + |\psi|^2 \psi = 0,
\end{eqnarray}
which is obtained from (\ref{dim_NLSE}) by neglecting the losses and using the rescaled variables $\psi = \Psi/\sqrt{P_0}$, $\tau = T\sqrt{\gamma P_0/ |\beta_2|}$, $z = Z\gamma P_0$, where $P_0$ is a typical value of the optical power $|\Psi|^2$. The IST analysis of the NLSE wavefields relies on the linear auxiliary Zakharov-Shabat (ZS) system for a vector wave function $\Phi = (\phi_1,\phi_2)^{\text{T}}$:
\begin{eqnarray}
\widehat{\mathcal{L}} \Phi = \lambda \Phi, \quad\quad\quad \widehat{\mathcal{L}} = 
\begin{pmatrix}
i \partial_{\tau}   & -i \psi(\tau) \\ -i \psi^*(\tau)   & - i \partial_{\tau}
\end{pmatrix},
\label{ZSsystem}
\end{eqnarray}
where $\lambda = \xi + i \eta$ is the $z$-independent complex spectral parameter and $\text{T}$ means transposition. The NLSE wavefield $\psi(\tau)$, playing the role of potential for the system (\ref{ZSsystem}), in general contains a finite number of solitons $N$ and a portion of incoherent nonlinear dispersive waves \cite{NovikovBook1984,AblowitzBook1981}. Each soliton in the wavefield corresponds to a discrete eigenvalue $\lambda_n = \xi_n + i \eta_n $ of the operator $\widehat{\mathcal{L}}$, where $n = 1,..., N$ and $\eta_n>0$. The eigenvalue defines asymptotic soliton  amplitude $A$ and velocity $V$ as,
\begin{eqnarray}
\label{An and Vn}
A_n = 2\eta_n, \quad\quad V_n = 2\xi_n.
\end{eqnarray}

\begin{figure}[!t]\centering
	\includegraphics[width=8.7cm]{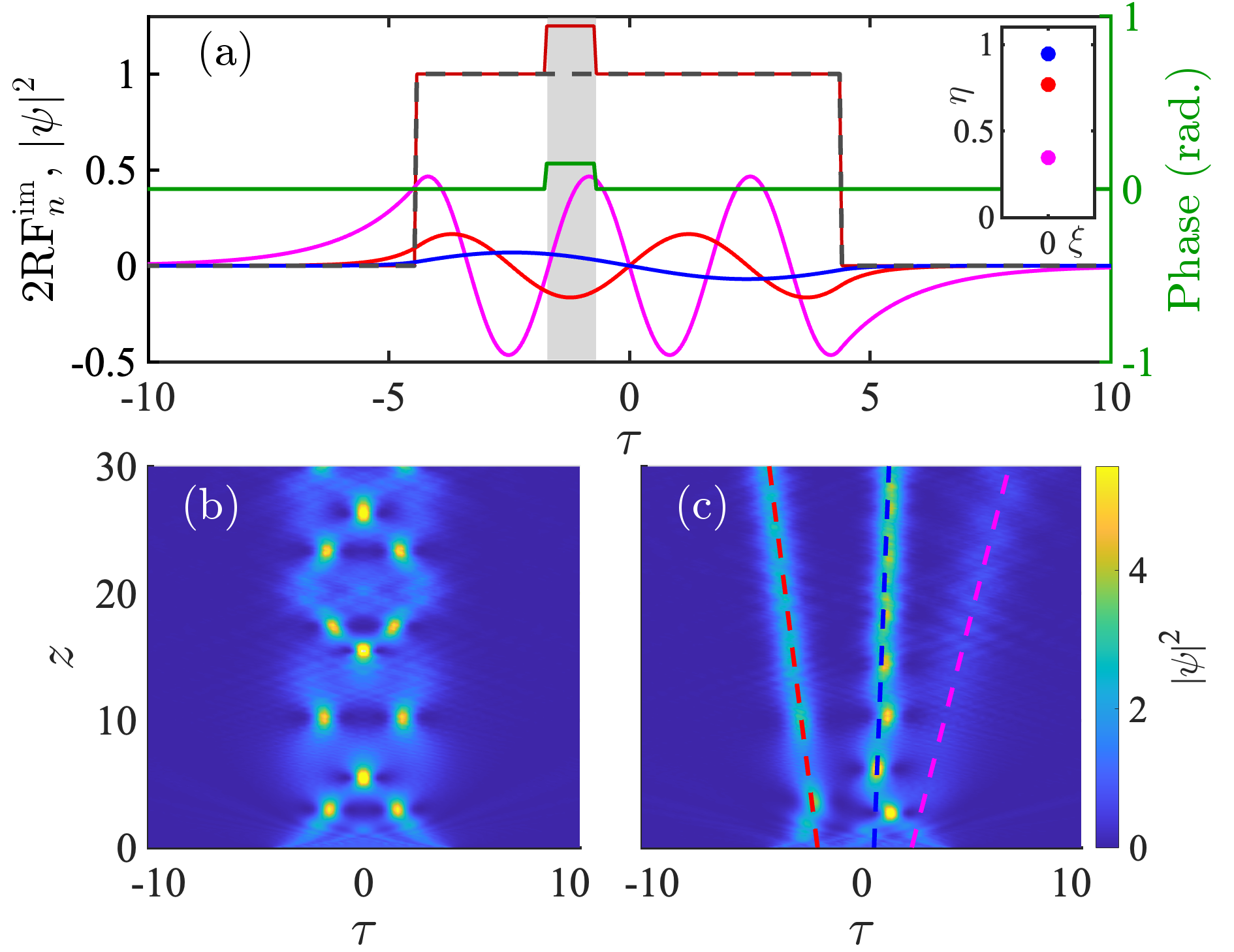}
	\caption{Eigenvalue response functions (RFs) of the box-shaped pulse with $L=8.85$ and manipulation of its solitons in numerical simulation. (a) The initial pulse (dashed gray) and $\text{RF}^{\text{im}}_{n}$. The pulse contains three solitons with the eigenvalues indicated on the inset and colored following the colors of the RFs. (b) Reference evolution of the unperturbed pulse in numerical simulation. (c) Evolution of the pulse having local imaginary perturbation; see red and green solid lines for power and phase modulation of the initial field in (a). Dashed lines in (c) show soliton trajectories predicted by theoretical expressions (\ref{dlambda_n}). The gray strip in (a) shows the area of the RF integration, which contributes to the soliton eigenvalue responses.}
\label{Fig1}
\end{figure}

When nonlinear dispersive waves are absent, the wavefield evolution obeys exact multi-soliton solutions, for $N=1$ given by the well-known single soliton formula,
\begin{eqnarray}
    \psi_{\text{s}} = A_1 \frac{\exp{\{ iV_1(\tau - \tau_{0}) + i (A_1^2 + V_1^2)z/2 +i\theta_0 \}}}{\cosh{A_1 (\tau - V_1 z - \tau_{0})}},
\end{eqnarray}
where $\tau_{0}$ and $\theta_0$ are soliton position and phase.

In the presence of nonlinear dispersive waves exact analytical solutions of the NLSE (\ref{NLSE}) are not available; however analysis of soliton eigenvalues allows one to access the main features of the nonlinear wavefield behavior. Consider a box-shaped real pulse $\psi_\sqcap$ of amplitude $A$ and width $L$, so that $\psi_\sqcap (\tau) = A$ for $|\tau| < L / 2$ and $\psi_\sqcap (\tau) = 0$ otherwise, see Fig.\,\ref{Fig1}(a). The pulse contains $N = \text{Integer}[ 1/2 + A L / \pi]$ zero velocity solitons with eigenvalues aligned on the imaginary axis: $\{ \lambda_n \} = i \{ \eta_n \}$ for $n=1,..,N$, where $\eta_n$ obey the following transcendental equation \cite{manakov1973nonlinear}:
\begin{eqnarray}
\label{azeros}
&& \tan (\chi_n L) = -  \chi_n/\eta_n, \;\;\;\;\; \chi_n = \sqrt{A^2 - \eta_n^2}.
\end{eqnarray}
For an arbitrary-shaped pulse, the eigenvalues can be found numerically using various available IST algorithms \cite{YangBook,BofOsb1992,wahls2015fast,vasylchenkova2018contour,mullyadzhanov2019direct}.

\begin{figure*}[!t]\centering
	\includegraphics[width=17.65cm]{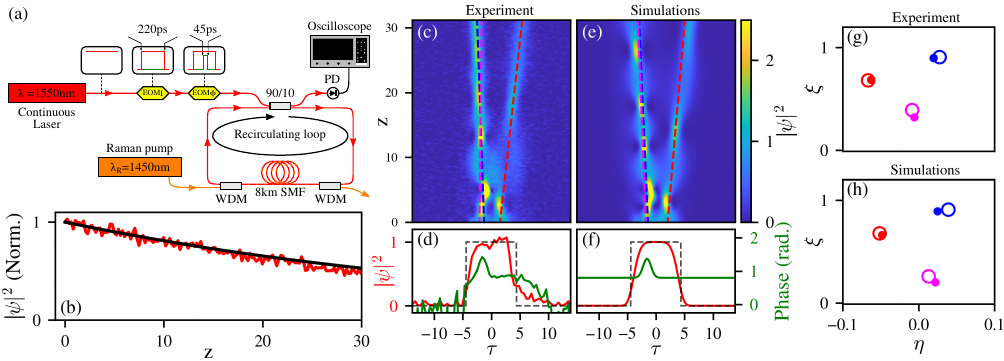}
	\caption{(a) Schematic experimental setup. (b) Measured power decay rate of the circulating light (red) and its exponential fit (black) with $\alpha = \qty{4d-7}{m^{-1}}$ in normalized units. (c, e) Spatiotemporal diagrams of the perturbed pulse propagation over $1200$ km, with soliton trajectories predicted by theoretical expressions (\ref{dlambda_n}). (d, f) Initial pulse power (red), its phase profile (green) and the approximating box-shaped potential used to compute the RFs. (g, h) Discrete IST spectrum: direct numerical computation (circles) and theoretical eigenvalues computed from expressions (\ref{dlambda_n}) (dots). Panels (c, d, g) demonstrate the experimental data, while (e, f, h) show corresponding numerical simulations of the full model [Eq.~(\ref{dim_NLSE})].}
\label{Fig2}
\end{figure*}

The spatiotemporal dynamics $\psi_\sqcap(z,\tau)$ features the oscillating interactions of $N$ solitons and the radiation, see Fig.\,\ref{Fig1}(b) obtained by numerical integration of the NLSE (\ref{NLSE}).  An instant disturbance of the pulse $\psi(\tau) \rightarrow \psi(\tau) + \delta\psi(\tau)$ leads to a change of the initial soliton eigenvalues, that for relatively weak $\delta\psi(\tau)$ can be captured by the first-order IST perturbation theory \cite{KaupSIAM1976, KarpmanJETP1977, KivsharRMP1989}:
\begin{eqnarray}
\delta\lambda = \frac{\langle \Phi^\dagger, \delta\widehat{\mathcal{L}}~\Phi \rangle}{\langle \Phi^\dagger, \Phi \rangle},
\quad
\delta\widehat{\mathcal{L}} = 
-i
\begin{pmatrix}
0     &    \delta\psi    \\     \delta\psi^*    &     0
\end{pmatrix}.
\label{dlambda}
\end{eqnarray}
Here $\Phi(\tau)$ represent solution of the ZS system with potential $\psi(\tau)$, $\Phi^\dagger = (\phi^*_2,\phi^*_1)^{\text{T}}$ and the scalar product $\langle f, g \rangle = \int_{-\infty}^{\infty} f^* g d\tau$.

For the rectangular potential $\psi_\sqcap (\tau)$ and its perturbation $\delta\psi(\tau)$ represented by a sum of Fourier harmonics, the integrals (\ref{dlambda}) were evaluated recently in \cite{mullyadzhanov2021solitons}. Here, to simplify the analysis of the problem (\ref{dlambda}), we introduce the eigenvalue response functions (RFs) for arbitrary real-valued $\delta\psi^{\text{re}}(\tau) = f (\tau)$ and imaginary-valued $\delta\psi^{\text{im}}(\tau) = if (\tau)$ pulse perturbations as,
\begin{eqnarray}
\label{dlambda_n}
    &&\delta\lambda_n = \int_{-\infty}^{\infty} \text{RF}(\tau)^{\text{re/im}}_n f(\tau) d\tau\,,
    \\\nonumber
    &&
    \text{RF}^{\text{re}}_n = -i \frac{\phi_1^2 + \phi_2^2}{\Delta} \biggl|_{\lambda=\lambda_n} 
    \,, \quad 
    \text{RF}^{\text{im}}_n = \frac{-\phi_1^2 + \phi_2^2}{\Delta} \biggl|_{\lambda=\lambda_n} \,,
    \\\nonumber
    &&
    \Delta = \langle \Phi^\dagger, \Phi \rangle =  2 \int_{-\infty}^{\infty} \phi_1 \phi_2 d\tau\,.
\end{eqnarray}

Eq.~(\ref{dlambda_n}) shows that the eigenvalue response to the pulse perturbation represents the integration of $f(\tau)$ with $\text{RF}(\tau)$. The latter plays the role of a weight function that provides an intuitively simple way to manipulate characteristics of individual solitons by managing $f(\tau)$. Remarkably, the functions $\text{RF}^{\text{re/im}}_n (\tau)$ are always either pure imaginary or pure real-valued; see Supplemental Material \cite{SuppM}, where we provide exact expressions for them. Thus, the real-valued perturbations of the box pulse can change only the imaginary part of $\lambda_n$, i.e., affect soliton amplitudes $A_n$, which is in line with general properties of the ZS system \cite{klaus2002purely}. Meanwhile, imaginary-valued perturbations act vice versa and change only soliton velocities $V_n$.

Here, we focus on the manipulation of soliton velocities as they can be easily detected in experiments by tracking soliton trajectories; see also the previous theoretical developments on the application of IST perturbation theory on the decay of soliton complexes \cite{satsuma1974b,prilepsky2007breakup}. Fig.\,\ref{Fig1}(a) shows typical behaviour of $\text{RF}(\tau)^{\text{im}}_{n}$ for a pulse containing three solitons affected by imaginary perturbation. The RFs being formed by discrete spectrum eigenfunctions of the ZS system exhibit an oscillatory behavior inside the box-shaped potential with zero value in its center. The number of the RFs maxima/minima increases with decreasing $\eta_n$, i.e., soliton amplitude. Importantly, for an unperturbed box-shaped pulse, solitons are not individualized due to their strong overlap and interaction. Nevertheless, the IST theory enables manipulation of individual solitons by applying localized perturbations.

\begin{figure*}[!t]\centering
	\includegraphics[width=17.65cm]{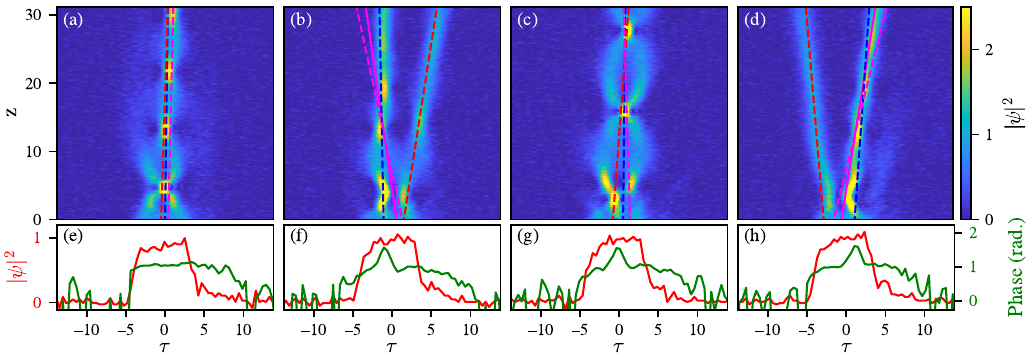}
	\caption{Experimental propagation of light pulses over $1200$ km: (a, e) no phase perturbation, (b-d, f-h) different locations of the phase perturbation. Panels structure is similar to Fig.\,\ref{Fig2}(c, d). Dashed lines show soliton trajectories obtained from theoretical expressions (\ref{dlambda_n}). In panels (b) and (d) we also added solid trajectories corresponding to the eigenvalues obtained from experimental IST of the perturbed pulse.}
\label{Fig3}
\end{figure*}

The soliton eigenvalue response is proportional to the RFs integral in the perturbation area; see the gray strip in Fig.\,\ref{Fig1}(a). This visual interpretation allows one to choose the areas of the pulse, where perturbations contribute the most, so one can selectively manipulate soliton velocities. As an example Fig.\,\ref{Fig1}(c) shows the pulse disintegration into three separate solitons moving along trajectoris well fitted by the theoretical formula (\ref{dlambda_n}) together with Eq.~(\ref{An and Vn}). The perturbed pulse exhibits dynamics of strongly interacting overlapped solitons, which asymptotically separate from each other. Note that, the perturbation is applied at a chosen position so that the two solitons with the smallest amplitudes receive nearly opposite significant velocities. In contrast, the soliton of the largest amplitude acquires a weakly positive velocity.

To verify the RFs approach we use the experimental setup represented schematically in Fig.\,\ref{Fig2}(a).
Here a continuous wave (CW) laser at \qty{1550}{nm} (APEX-AP3350A) is modulated in both intensity and phase by two electro-optic modulators EOM$_{I - \phi}$ (iXblue MX-LN-20 and iXblue MPZ-LN-10). The setup enables the design of various initial conditions consisting of square-shaped pulses with superimposed localized phase perturbations. The phase perturbation of the field $\psi \rightarrow \psi e^{i\phi}\simeq \psi(1+i\phi)$, allows us to generate almost imaginary perturbations triggering the velocity of solitons [Eq.~(\ref{dlambda_n})]. The designed initial conditions are launched into a recirculating fiber loop consisting of an \qty{8}{km}-long  single-mode fiber (SMF) closed back on itself with a $90/10$ fiber coupler. Light circulating in the loop is partially extracted at each round trip through the \qty{10}{\%} arm of the fiber coupler and detected by a calibrated fast photodiode (PD) (Finisar - XPDV2120R-10-20-040) coupled to an \qty{80}{GSa/s} sampling oscilloscope (TeledyneLecroy - LabMaster 10-Zi \qty{65}{GHz}) leading to a detection bandwidth around \qty{36}{GHz}. Additionally, the losses accumulated over one circulation in the fiber loop are partially compensated by a counter-propagating Raman pump coupled in and out of the loop via wavelength division multiplexers (WDM), reducing the effective power decay rate of the circulating field to \qty{0.0017}{dB/km} or $\alpha = \qty{4d-7}{m^{-1}}$, see Fig.~\ref{Fig2}(b). Our data analysis enables the reconstruction of the spatiotemporal dynamics of the intensity recorded in single-shot  \cite{kraych2019nonlinear,Kraych2019,suret2023soliton}. Additionally, the phase dynamics of the initial conditions is measured by using a $3\times 3$ coupler-based interferometric technique; see \cite{kelleher2010phasor} and Supplemental Material \cite{SuppM}. The full characterization of the initial light field in terms of phase and amplitude allows us to compute the experimental IST spectrum.

In the experiments, the propagation of square-shaped pulses with $P_0 = \qty{20}{mW}$ and full width at half maximum (FWHM) of \qty{220}{ps} perturbed by a Gaussian-shaped localized phase modulation ($\text{FWHM}\,\sim\,45\text{ps}$) was recorded over a distance of \qty{1200}{km}. Figs.\,2(c,d) show $150$ round trips inside the fiber loop, that corresponds to the propagation of a pulse with $\text{FWHM}=8.75$ over 30 nonlinear lengths in normalized units. The generated pulse is not a sharp box due to the limited bandwidth of EOMs; however it is well approximated by a super-Gaussian function, which is close to our theoretical model, see Fig.\,2(d,f) and Supplemental Material \cite{SuppM}. The spatiotemporal dynamics recorded in the experiment [Fig.\,2(c)] is very similar to the one obtained in numerical simulations of the full NLSE model (\ref{dim_NLSE}) [Fig.\,3(e)] and demonstrates the disintegration of the pulse induced by the phase perturbation. The discrete IST spectra of the perturbed pulse measured in the experiment and its counterpart from simulations are plotted with circles in Figs.~2(g,h).

Theoretical description of the experiments starts from computing perturbations to the soliton eigenvalues of the initial pulse $\psi(\tau)$, using a box potential of the same power value $\int |\psi|^2 d\tau$, see Fig.~2(d). To account for the imperfections of the box-shaped light generation, we use experimentally measured soliton eigenvalues of the unperturbed pulse as the initial IST spectrum, to which we add the theoretical corrections in accordance with the applied phase perturbation; see Supplemental Material \cite{SuppM} for details of the data analysis procedures. Our theory predicts that the largest and smallest solitons acquire similar negative velocities. In contrast, the middle-amplitude soliton acquires a positive velocity, so the perturbed pulse decays into two parts. Experimental results match very well our theoretical predictions; see Fig.~2(c) for the spatiotemporal diagram of the light propagation and the theoretical trajectories. In particular, the left detached pulse exhibits strong internal oscillations produced by the interaction of two solitons moving together. In addition, we investigate experimentally the influence of the perturbation location on the velocities of individual solitons [Fig.~3], which includes a reference example of an unperturbed pulse [Figs.~3(a)]. The latter demonstrates the standard oscillating dynamics similar to the theoretical one shown in Fig.~1(b), moderately affected by the power loss. Fig.~3(b) shows an analog of the experiment from Fig.~2(c) in the case when the perturbation is closer to the pulse center. Meanwhile, Fig.~3(d) is another scenario when the middle-amplitude soliton detaches to the left while the remaining two propagate together to the right. As expected, perturbation located at the pulse center does not disintegrate the light pulse since the functions $\text{RF}^{\text{im}}_{n}$ are anti-symmetric and thus, solitons do not acquire velocities, see Fig.~3(c). Note that in Figs.~3(b,d), the smallest amplitude solitons exhibit moderate discrepancy with theory, which can be mainly attributed to second-order corrections of the perturbation theory. We confirm this conjecture by presenting theoretical trajectories computed from experimental IST eigenvalues, see solid lines in Figs.~3(b,d), which considerably better fit the experimentally observed data.

Our approach highlights that the wave functions of the ZS system play not only an auxiliary role in the IST construction but also are directly linked to the physical properties of the nonlinear wave fields governed by the NLSE. Like the probability amplitude in quantum mechanics formalism \cite{landau1958quantum}, the constructed RFs tell us how perturbations address certain solitons even though the latter are completely delocalized within the pulse due to strong overlapping and interaction. Further studies of the wave function properties will, we believe, illuminate the description of complex scenarios of the NLSE dynamics \cite{akhmediev2009extreme,onorato2013rogue,el2016dam,conforti2018auto,bendahmane2022piston}, benefit to the perspective IST-based optical telecommunications approaches \cite{le2017nonlinear,turitsyn2017nonlinear,frumin2017new} and contribute to the rapidly developing field of integrable turbulence and dense soliton gases \cite{agafontsev2015integrable,walczak2015optical,pelinovsky2013two,soto2016integrable,Kraych2019,redor2019experimental,slunyaev2022statistical,suret2024soliton}. The proposed here RFs method of examining nonlinear pulses sensitivity is very general and can be applied to other integrable systems, such as the Korteweg–de Vries or sine-Gordon equations \cite{NovikovBook1984}, as well as other wave shapes, e.g., to the $\text{sech}$-pulse, when the wave functions are known \cite{satsuma1974b}. When wave function solutions are unavailable, the RFs can be evaluated numerically and analyzed graphically, extending the capacity of the IST perturbation theory \cite{kaup1976perturbation,karpman1977perturbation,KivsharRMP1989,YangBook} along with the recently advanced semi-analytical IST approaches \cite{gelash2024bi}. Putting forward experimental techniques for control and manipulation of solitons in optical fibers \cite{jang2015temporal,suret2023soliton,englebert2024manipulation}, we emphasize that they can be also applied in nearly integrable hydrodynamical or Bose-Einstein condensate systems, see e.g. \cite{chabchoub2013hydrodynamic,redor2019experimental,mossman2024observation}.

\onecolumngrid

\section{Supplementary materials}

\subsection{Eigenvalue Response Functions}

Here we derive complete theoretical expressions for the eigenvalue response functions (RFs) of the box-shaped potentials. 
We employ the corresponding wave functions $\Phi(\tau, \lambda) = (\phi_1,\phi_2)^\mathrm{T}$ of the Zakharov-Shabat (ZS) system. 
For convenience, we write the latter in the following form, which is equivalent to the Eq.~(3) of the main Letter,
\begin{eqnarray}
	\Phi_{\tau} &=& \begin{pmatrix}
	-i \lambda & \psi(\tau) \\ -\psi^*(\tau) & i \lambda
	\end{pmatrix}
	\Phi.
	\label{ZSsystem1}
\end{eqnarray}

For the box-shaped potential, $\psi_\sqcap (\tau)$ of length $L$ and amplitude $A$, the wave function $\Phi(\tau)$ can be obtained by solving the scattering problem outside and inside the box and matching the solution at $\tau = \pm L/2$, see Ref. [26] of the main Letter, leading to the following result at an arbitrary spectral parameter $\lambda$:
\begin{eqnarray}
\label{phi1}
\phi_1 &=& 
    \begin{cases}
      e^{- i \lambda \tau}, & \text{if $\tau<-L/2$}\\
      \{-i\lambda\sin (s) \chi^{-1} + \cos (s)\} e^{i L \lambda/2}, & \text{if $-L/2<\tau<L/2$} \\
      a(\lambda) e^{- i \lambda \tau}, & \text{if $\tau>L/2$}
    \end{cases} 
\\
\label{phi2}
\phi_2 &=& 
    \begin{cases}
      0, & \text{if $\tau<-L/2$} \\
      -A \chi^{-1} \sin(s) e^{i L \lambda/2}, & \text{if $-L/2<\tau<L/2$} \\
      b(\lambda) e^{i \lambda \tau}, & \text{if $\tau>L/2$}
    \end{cases}  
\end{eqnarray}
where
\begin{eqnarray}
    \chi = \sqrt{A^2 + \lambda^2}, \qquad 
    s = (L/2+\tau)\chi,
\end{eqnarray}
while $a(\lambda)$ and $b(\lambda)$ are the so-called scattering coefficients, see Ref. [14-15] of the main Letter for details on the IST theory formalism,
\begin{eqnarray}
    \label{coeff_a}
    a(\lambda) &=& e^{iL\lambda} \bigl\{ \cos(\chi L) - i\lambda \sin(\chi L)\chi^{-1} \bigr\},
    \\
    b(\lambda) &=& -A\sin(\chi L)\chi^{-1}.
\end{eqnarray}

According to Eq.~(8) of the main Letter, the RFs corresponding to real-valued and imaginary-valued perturbations of the box-shaped pulse evaluated at the discrete eigenvalue points $\lambda = \lambda_n$ are given by,
\begin{eqnarray}
\label{RFfinal}
    \text{RF}^{\text{re}}_n (\tau) =
    -\frac{i}{2}\frac{\phi_1^2 + \phi_2^2}
    {\int_{-\infty}^{\infty} \phi_1 \phi_2 d\tau}\biggl|_{\lambda=\lambda_n}\,,
    \qquad\qquad 
    \text{RF}^{\text{im}}_n (\tau) =
    -\frac{1}{2}\frac{\phi_1^2 - \phi_2^2}
    {\int_{-\infty}^{\infty} \phi_1 \phi_2 d\tau}\biggl|_{\lambda=\lambda_n}\,.
\end{eqnarray}
As one can see, from the structure of  the wave function components $\phi_1(\tau)$ and $\phi_2(\tau)$ are provided above in Eqs.~(\ref{phi1}), (\ref{phi2}) the functions $\text{RF}^{\text{re}}_n (\tau)$ and $\text{RF}^{\text{im}}_n (\tau)$ are always pure imaginary-valued and pure real-valued. 
Note that according to the IST theory, discrete eigenvalues are zeros for the scattering coefficient (\ref{coeff_a}), i.e. $a(\lambda)|_{\lambda = \lambda_n} = 0$, that can be also seen from the eigenvalue relation (6) provided in the main Letter. 
In addition, the second scattering coefficient $b(\lambda)|_{\lambda = \lambda_n} = (-1)^n$. 
Evaluating integrals in expressions (\ref{RFfinal}) at $\lambda_n = i\eta_n$, and after some transformations we obtain the following closed-form expressions for the box-shaped potential response functions,
\begin{eqnarray}
\label{RFre}
\text{RF}^{\text{re}}_n (\tau) &=& 
    -\frac{i}{\Delta_n}\begin{cases}
      e^{2\eta_n\tau}, & \text{if $\tau<-L/2$}\\
      \bigl\{ \bigl (\cos [s_n(\tau)] +\eta_n\chi^{-1}_n \sin [s_n(\tau)] \bigr )^2 + A^2 \chi^{-2}_n \sin^2 [s_n(\tau)] \bigr\} e^{-\eta_n L}, & \text{if $-L/2<\tau<L/2$} \\
      e^{-2\eta_n\tau}, & \text{if $\tau>L/2$}
    \end{cases} 
\\
\label{RFim}
\text{RF}^{\text{im}}_n (\tau) &=& 
    -\frac{1}{\Delta_n}\begin{cases}
      e^{2\eta_n\tau}, & \text{if $\tau<-L/2$}\\
       \bigl\{ \bigl (\cos [s_n(\tau)] +\eta_n\chi^{-1}_n \sin [s_n(\tau)] \bigr )^2 - A^2 \chi^{-2}_n \sin^2 [s_n(\tau)] \bigr\} e^{-\eta_n L}, & \text{if $-L/2<\tau<L/2$} \\
      -e^{-2\eta_n\tau}, & \text{if $\tau>L/2$}
    \end{cases} 
\end{eqnarray}
where $\chi_n  = \chi (\lambda_n)$, $s_n(\tau) = (L/2+\tau)\chi_n$, and the normalization constant coefficient $\Delta_n = 2\int_{-\infty}^{\infty} \phi_1 \phi_2 d\tau |_{\lambda=\lambda_n}$ is,
\begin{eqnarray}
    \Delta_n = -A\chi_n^{-2} e^{-\eta_n L} \bigl [ \sin^2 L\chi_n - \frac12 \eta_n \chi^{-1}_n \sin 2 L\chi_n + \eta_n L \bigr ].
\end{eqnarray}

\subsection{Numerical methods}\label{section:}

To locate soliton eigenvalues of the box-shaped potentials [Fig.~1(a) and Fig.~2(h) of the main Letter], we solve numerically the transcendental Eq.~(6) of the main Letter at fixed parameters $A$ and $L$. In the case of experimental wave fields [Fig.~2(g) of the main Letter] we use the standard Fourier collocation method, allowing one to solve the ZS eigenvalue problem numerically; see Ref. [22] of the main Letter for details on its implementation. Being fast and fairly accurate, the Fourier collocation approach is based on the Fourier decomposition of the wave field and transformation of the ZS differential equations [Eq.~(\ref{ZSsystem1})] into a matrix algebraic equations. The latter provides a straightforward access to the ZS system eigenvalues. Alternatively, soliton eigenvalues can be found using more advanced IST algorithms, see Refs. [27-30] in the main Letter.

The conservative NLSE evolution presented in Fig.~1(b,c) of the main Letter was simulated using standard pseudospectral second-order split-step (Fourier)  method. The pulse discretization step was $\Delta \tau = 0.04$ providing accurate description of the wavefield Fourier spectra, while 
the propagation step $\Delta z$ was chosen in order to satisfy the algorithm stability criteria $\Delta z < \Delta \tau^2/4$. For the NLSE evolution with losses, presented in Fig.~2(e) of the main Letter, the same method was used with the addition of an exponential decay term, $\exp(-\alpha\Delta z/2)$ , affecting the wavefield as it numerically propagates, with $\alpha = \qty{4d-7}{m^{-1}}$ or $\alpha = 0.016$ in normalized units. 
The discretization and propagation steps for the pulse shown in Fig.~2(f) of the main Letter were chosen similar to the conservative NLSE case.


\subsection{Phase measurement using $3\times 3$ coupler}

\begin{figure*}[!h]\centering
	\includegraphics[width=18cm]{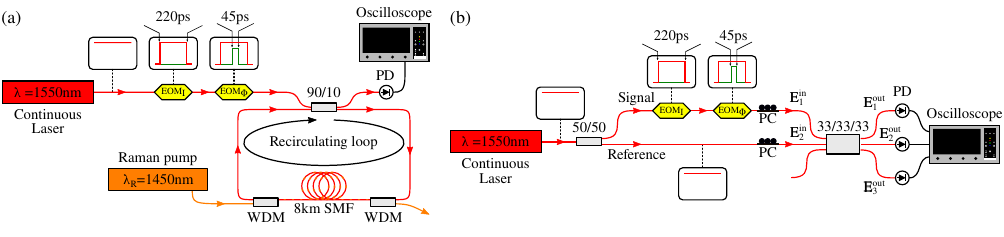}
	\caption{(a) Schematic experimental recirculating fiber loop setup. (b) Experimental setup for the phase measurement of the initial condition. EOM$_\text{I}$ : Intensity electro-optic modulator. EOM$_{\phi}$ : Phase electro-optic modulator. WDM : Wavelength-division multiplexing. SMF : Single mode fiber. PD : Photodiode. PC : Polarization controller.}
\label{Drawing2}
\end{figure*}

The ideal $3\times 3$ coupler distributes the injected intensity equally across each of the input ports to each of the output ports. We denote the electric field at the output or input port $j$ as $E^{\text{out}}_j$ and $E^{in}_j$. Each output consists of its own respective input and the phase-shifted contribution of $2\pi/3$ from the other two inputs. For example, the output from port $1$ is given by:

\begin{equation}
E^{\text{out}}_1 = \frac{1}{\sqrt{3}} \left[ E^{\text{in}}_1 + E^{\text{in}}_2e^{i2\pi/3} + E^{\text{in}}_3e^{i2\pi/3} \right].
\label{syst_3x3}
\end{equation}

In our setup, port $1$ is used for the signal, port $2$ for the reference, and port $3$ is unused, simplifying the system of $E^{\text{out}}_j$. Considering the relative phase $\Delta\phi$ associated with the signal, we obtain the relations for the intensities:

\begin{equation}
\left\{
\begin{array}{lll}
I^{\text{out}}_1 = \frac{1}{3} \left[ I^{\text{in}}_1 + I^{in}_2 + 2\sqrt{I^{\text{in}}_1I^{\text{in}}_2} \cos(\Delta\phi-2\pi/3) \right], \\
\\
I^{\text{out}}_2= \frac{1}{3} \left[ I^{\text{in}}_1 + I^{in}_2 + 2\sqrt{I^{\text{in}}_1I^{\text{in}}_2} \cos(\Delta\phi+2\pi/3) \right], \\
\\
I^{\text{out}}_3 = \frac{1}{3} \left[ I^{\text{in}}_1 + I^{\text{in}}_2 + 2\sqrt{I^{\text{in}}_1I^{\text{in}}_2} \cos(\Delta\phi) \right].
\end{array}
\right.
\end{equation}

Combining these expressions we obtain the following relation for the phase in the signal's frame of reference,

\begin{equation}
\tan(\Delta\phi) = \frac{\sqrt{3}}{2} \left[\frac{I^{\text{out}}_1 - I^{\text{out}}_2}{I^{\text{out}}_3 - I^{\text{out}}_1/2 - I^{\text{out}}_2/2}\right].
\label{Eq 3x3}
\end{equation}

The phase term appears in the tangent function, which is periodic and has discontinuities at $\pm\pi/2$, leading to function jumps when the phase changes monotonically. Using Eq.~(\ref{Eq 3x3}) we find expressions for $\cos(\Delta\phi)$ and $\sin(\Delta\phi)$,

\begin{equation}
    \begin{aligned}
        \sin(\Delta\phi) = \frac{\sqrt{3}}{2} \left[ \frac{ I^{\text{out}}_1 - I^{\text{out}}_2}{ \sqrt{I^{\text{in}}_1I^{\text{in}}_2}}\right]
        \,,\qquad 
        \cos(\Delta\phi) = \frac{I^{\text{out}}_3 - I^{\text{out}}_1/2 - I^{\text{out}}_2/2}{ \sqrt{I^{\text{in}}_1I^{\text{in}}_2}}\,.
    \end{aligned}
\end{equation}

Now, with the expressions for $\cos(\Delta\phi)$, $\sin(\Delta\phi)$ and $\tan(\Delta\phi)$, we impose conditions on the relative phase measurements
and unfold the phase value across the entire range of the pulse amplitudes.

\begin{figure*}[!h]\centering
	\includegraphics[]{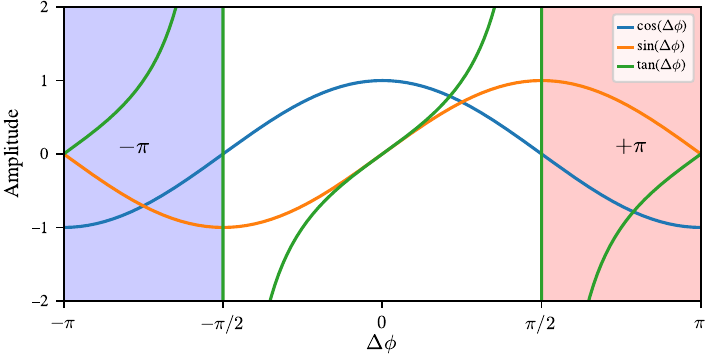}
	\caption{Behaviour of the cosine, sine, and tangent functions in the three colored zones, separated by the discontinuities of the tangent function associated with the phase unwrapping conditions (\ref{Eq 3x3}). To enable unambiguous finding of the light pulse phase modulations we add $\pm\pi$ to the phase measurement in the left (blue) and right (red) zones.}
\label{Fig10}
\end{figure*}

In Fig.\,\ref{Fig10} we represent the cosine, sine, and tangent functions over the interval $[-\pi;\pi]$ with three colored zones separated by the discontinuities of the tangent function at $\pm\pi/2$. In the central zone (white), the cosine always has a positive value, while in the left (blue) and right (red) zones, it is negative. The sine, on the other hand, is positive from $[0,\pi]$ and negative from $[-\pi,0]$. This representation allows us to determine the phase value implementing the following procedure,
\begin{itemize}
\item Left (blue) zone: adding $\pi$ to the phase measurement.
\item Central (white) zone: no phase measurement corrections.
\item Right (red) zone: subtracting $\pi$ from the phase measurement.
\end{itemize}
Using this procedure, we avoid phase jumps, enabling an unambiguous measurement (up to a general constant factor) of the whole light pulse phase modulations.

\subsection{Experimental data processing}

\begin{figure*}[!h]
	\includegraphics[width=13.4cm]{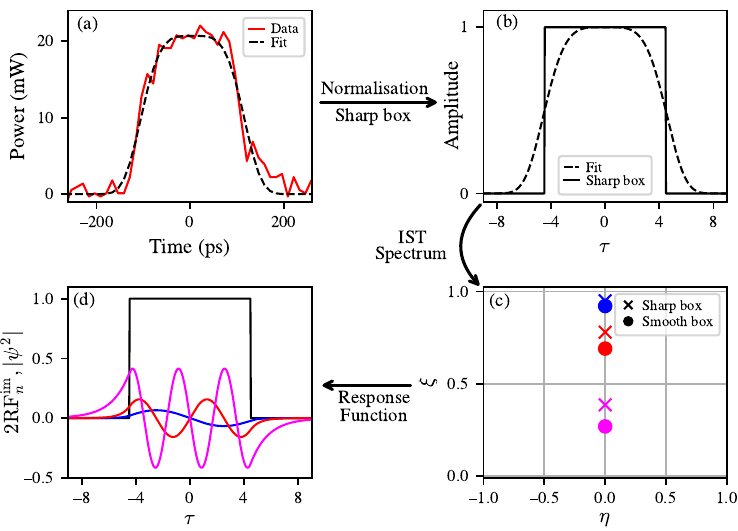}
	\caption{Experimental data processing protocol. (a) In red line: initial light pulse (at $z = 0\,\text{km}$). In black dotted line: the super-Gaussian fit with (\ref{Fit smooth}). (b) In black dotted line: the super-Gaussian fit potential obtained from the fit with $P = 2.5$. In black line: the equivalent box-shaped potential. (c) IST spectrum: dots for the super-Gaussian potential (smooth box), crosses for the box-shaped potential. (d) Eigenvalue response functions (RFs) for imaginary perturbations associated to each soliton of the box-shaped potential. The color scheme is the same as used for the IST spectrum in (c).}
\label{Fig5}
\end{figure*}

Fig.\,\ref{Fig5} illustrates our data processing protocol applied to each initial experimental pulse.
The protocol enables the characterization of the initial condition in terms of its discrete IST spectrum and the eigenvalue response functions (RFs).
In the first step, see Fig.\,\ref{Fig5}(a), we fit the signal with a super-Gaussian function $f(T)$, extracting the peak power of our pulses $P_0$,
which we use for normalizing the experimental data,
\begin{equation}
    f(T) = P_0\exp\left(-\left(\frac{(T-T_0)^2}{2\sigma^2}\right)\right)^P
    \label{Fit smooth}
\end{equation}
In the second step, see Fig.\,\ref{Fig5}(b); since the sensitivity functions are defined for a sharp box-shaped potential, we find an equivalent box-shaped version of our experimental super-Gaussian fit. We use the equality of the potential surfaces as a criterion because it preserves the number of solitons, $N = \text{Integer}[ 1/2 + A L / \pi]$. For the pulse shown in Fig.\,\ref{Fig5}, we obtain the length of the equivalent box $L_{\text{box}} = 8.75$.

Then, we compute the IST spectrum of the super-Gaussian and box fits numerically and compare their soliton composition as shown in Fig.\,\ref{Fig5}(c). Both potentials have three solitons with no relative velocity. The amplitudes of the super-Gaussian fit potential solitons are slightly lower than their box potential counterpart. This feature is more pronounced for the smallest soliton as it is more sensitive to perturbations. Finally, using the eigenvalues from the box-shaped potential, we retrieve the eigenvalue RFs, which we use for theoretical predictions of soliton trajectories in the main Letter, see Fig.\,\ref{Fig5}(d).

\subsection{Soliton trajectories}

In our experiments we perturb the initial field $\psi(\tau)$ by phase modulation $\phi(\tau)$. The modulation is characterized by a Gaussian shape with a full width at half maximum (FWHM) of $\qty{48}{ps} \approx \qty{1.65}{\tau}$ and an average amplitude of \qty{0.55}{rad}. We perform the phase measurement independently from the recording of the spatio-temporal dynamics. Then we combine two separate recordings and reconstruct the initial complex field. Finally, we obtain the perturbation term $\delta\psi(\tau)$ as, 
\begin{equation}
    \delta\psi(\tau)  = \psi(\tau)_{\text{pt}} - \psi(\tau) = \psi(\tau) e^{i\phi(\tau)} - \psi(\tau),
\end{equation}
and complete the data needed for the prediction of soliton trajectories.

The perturbation $\delta\psi(\tau)$ is primarily imaginary-valued for small-amplitude phase modulations $\phi(\tau)$; however, it also has a real-valued component that we account for when calculating the deviation of the eigenvalues. We obtain two types of the IST spectrum -- theoretical and experimental, shown in Fig.~2(g,h) of the main Letter by dots and circles. The dots (theory) correspond to the following two-step procedure: i) first we compute initial IST spectrum for absolute-valued intensity profiles leading to eigenvalues alight along imaginary axis, ii) then with the eigenvalue response functions we predict theoretically the soliton eigenvalue deviations corresponding to the perturbation $\delta\psi(\tau)$ and add them to the initial IST spectrum. The circles (experiment) are experimental soliton eigenvalues, obtained directly from the complex-valued field $\psi(\tau)_{\text{pt}}$.

\begin{figure*}[!h]
	\includegraphics[width=11.5cm]{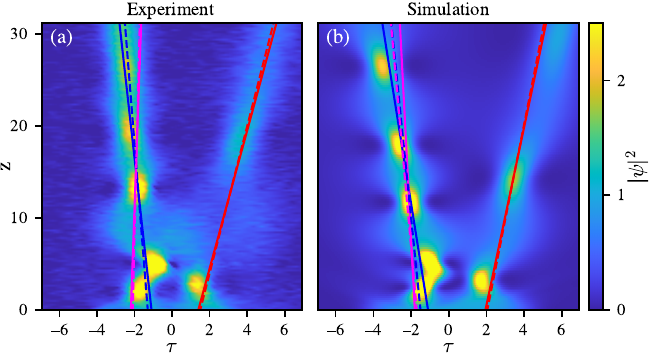}
	\caption{Spatiotemporal diagrams of the perturbed pulse propagation from Fig.~2(c,e) of the main Letter. In addition to theoretical trajectories of soliton propagation (dashed lines), solid lines show the trajectories obtained from direct numerical computation of the IST spectrum.}
\label{Fig111}
\end{figure*}

Once the real parts of the perturbed soliton eigenvalues is found, we compute soliton velocities (see Eq.~(4) of the main Letter) and 
plot soliton trajectories. The incline of the trajectories corresponds to soliton velocities, while their origin is chosen so to fit position of solitons at large distance where they well separated ($z\gtrapprox 15$).

\section{Acknowledgments}
This work of PS, FC, AM, SR has been partially supported by the Agence Nationale de la Recherche through the LABEX CEMPI project (ANR-11-LABX-0007), the SOGOOD (SOGOOD ANR-21-CE30-0061) and StormWave (StormWave ANR-21-CE30-0009) projects, the Ministry of Higher Education and Research, Hauts de France council and European Regional Development Fund (ERDF) through the Nord-Pas de Calais Regional Research Council and the European Regional Development Fund (ERDF) through the Contrat de Projets Etat-Région (CPER Wavetech). PS, FC, AM and SR thank the Centre d'Etudes et de Recherches Lasers et Applications CERLA for the technical help and and equipment. 
The authors would like to thank the Isaac Newton Institute for Mathematical Sciences, Cambridge, for its support and hospitality during the program "Emergent phenomena in nonlinear dispersive waves," during which significant progress was made on this work.
The work of RM on the response functions derivation was supported by RSF Grant No. 19-79-30075-$\Pi$.
The work of AG was funded by the European Union's Horizon 2020 research and innovation program under the Marie Skłodowska-Curie grant agreement No. 101033047.
%

%

\end{document}